\documentclass[12pt]{article}
\usepackage{epsfig}
\begin{document}
%
\def\be{\begin{equation}}
\def\ee{\end{equation}}
\def\bq{\begin{equation}}
\def\eq{\end{equation}}
\def\bqa{\begin{eqnarray}}
\def\eqa{\end{eqnarray}}
\def\roughly#1{\mathrel{\raise.3ex
\hbox{$#1$\kern-.75em\lower1ex\hbox{$\sim$}}}}
\def\lsim{\roughly<}
\def\gsim{\roughly>}
\def\llgm{\left\lgroup\matrix}
\def\rrgm{\right\rgroup}
\def\vectrl #1{\buildrel\leftrightarrow \over #1}
\def\partrl{\vectrl{\partial}}
\def\gslash#1{\slash\hspace*{-0.20cm}#1}



\begin{center}
{\bf On Ultrahigh-energy Neutrino Scattering}
\end{center}
 \vspace {0.5 cm}
\begin{center}
{\bf  Masaaki Kuroda}\\[2.5mm]
Institute of Physics, Meijigakuin University\\ [1.2mm] 
Yokohama, Japan\\ [3mm] 
{\bf Dieter Schildknecht} \\[2.5mm]
Fakult\"{a}t f\"{u}r Physik, Universit\"{a}t Bielefeld \\[1.2mm] 
D-33501 Bielefeld, Germany \\[1.2mm]
and \\[1.2mm]
Max-Planck Institute f\"ur Physik (Werner-Heisenberg-Institut),\\[1.2mm]
F\"ohringer Ring 6, D-80805, M\"unchen, Germany
\end{center}

\vspace{2 cm}

\baselineskip 20pt

\begin{center}
{\bf Abstract}
\end{center}
We predict the neutrino-nucleon cross section at ultrahigh energies
relevant in connection with the search for high-energy cosmic neutrinos.
Our investigation, employing the color-dipole picture, 
among other things allows us to
quantitatively determine which fraction of the ultrahigh-energy 
neutrino-nucleon cross section stems from the saturation versus the
color-transparency region. We disagree with various results 
in the literature that predict a strong suppression of 
the neutrino-nucleon cross section at neutrino energies above
$E \cong 10^9 GeV$. Suppression in the sense of a diminished increase
of the neutrino-nucleon cross section with energy only starts to occur
at neutrino energies beyond $E \cong 10^{14} GeV$.

\vfill\eject
Initiated by the experimental search for cosmic neutrinos of energies larger
than $E \simeq 10^6 {\rm GeV}$\footnote{Compare refs. 16-24 in \cite{a}}, 
the theoretical investigation\footnote{Compare e.g. refs. 2-8 in \cite{i}} 
of the neutrino-nucleon
interaction at ultrahigh energies received much attention recently. Predictions 
require a considerable extension of the theory of neutrino-nucleon deep
inelastic scattering (DIS) into a kinematic domain beyond the one where results
from experimental tests are available at present. Different theoretical
approaches have been employed ranging from conventional linear evolution of 
nucleon parton distributions to the investigation of possible non-linear
effects conjectured to becoming relevant in the ultrahigh-energy domain.

In the present note, we consider neutrino scattering in the framework of the
color dipole picture (CDP)\footnote{Compare ref. \cite{Lanz1} for recent
reviews on the CDP and an extensive list of references.}. 
The CDP is uniquely suited for a treatment of
ultrahigh-energy neutrino scattering. Extrapolating the results from 
electron-proton scattering at HERA, we expect the total neutrino-nucleon cross
section at ultrahigh energies to be dominantly due to the kinematic range of
$x \ll 0.1$ of the Bjorken variable $x_{bj} \equiv x \cong Q^2/W^2$. This
is the domain of validity of the CDP.

In particular, we shall focus on the question of color transparency versus
saturation. Does the total neutrino-nucleon cross section at ultrahigh
energies dominantly originate from the region of large values of the low-x
scaling variable \cite{Lanz2, 1108},

\be
\eta (W^2,Q^2) = \frac{(Q^2 + m^2_0)}{\Lambda^2_{sat} (W^2)},
\label{1}
\ee
namely $\eta (W^2,Q^2) \gg 1$ (``color transparency'' region), or is there a
substantial part that is due to the kinematic range of $\eta(W^2,Q^2)\ll 1$
(``saturation'' region)? 

In (\ref{1}), $\Lambda^2_{sat} (W^2)$
denotes the
``saturation scale'' that increases with the $\gamma^*(Z^0, W^\pm) p$
center-of-mass energy squared, $W^2$, as $(W^2)^{C_2}$, where 
$C_2 \simeq
0.29$ (compare (\ref{12}) below). 
At HERA energies, $\Lambda^2_{sat} (W^2)$ approximately ranges from
$2 {\rm GeV}^2 \lsim \Lambda^2_{sat} (W^2) \lsim 7 {\rm GeV}^2$. The 
$\gamma^* (Z^0, W^\pm)$ virtual four-momenta squared in (\ref{1}) is
denoted by $q^2 = - Q^2$, and $m^2_0 \simeq 0.15 {\rm
    GeV}^2$ 
(for light quarks).
Compare Fig. 1 for the $(Q^2,W^2)$ plane with the
line of $\eta (W^2,Q^2) = 1$.

\begin{figure}[h]
\begin{center}
\includegraphics{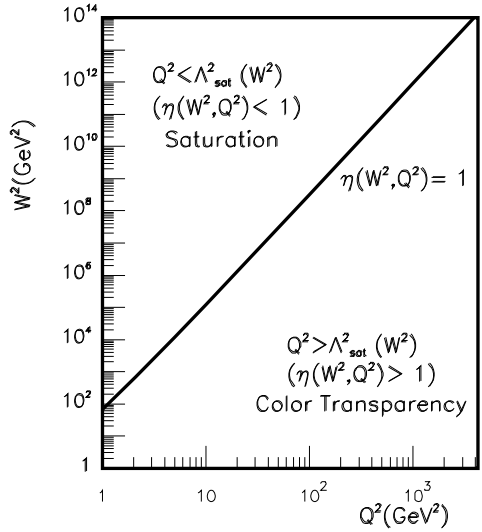}
\vspace*{-1cm}
 \caption{The $(Q^2,W^2)$ plane showing the line $\eta (W^2,Q^2) = 1$ that
separates the saturation region from the color-transparency region.}
\end{center}
\end{figure}

The charged-current neutrino-nucleon cross section we shall concentrate on,
as a function of the neutrino energy, $E$, is given by (e.g. \cite{b})
\be
\sigma_{\nu N} (E) = \int^s_{Q^2_{min}} dQ^2 \int^1_{\frac{Q^2}{s}} dx
\frac{1}{xs} \frac{\partial^2 \sigma}{\partial x \partial y},
\label{2}
\ee
where
\be
\frac{\partial^2 \sigma}{\partial x \partial y} = G^2_F \frac{s}{2 \pi}
\left( \frac{M^2_W}{Q^2 + M^2_W} \right)^2 \sigma_r (x, Q^2),
\label{3}
\ee
and $\sigma_r (x,Q^2)$ in (\ref{3}) denotes the ``reduced cross section''
\be
\sigma_r (x,Q^2) = \frac{1+(1-y)^2}{2} F_2^\nu (x,Q^2) - \frac{y^2}{2}
F_L^\nu (x,Q^2) + y (1-\frac{y}{2}) x F_3^\nu (x,Q^2).
\label{4}
\ee
In standard notation, $s$ denotes the neutrino-nucleon center-of-mass
energy squared,
\be
s = 2 M_pE + M_p^2 \cong 2 M_pE,
\label{5}
\ee
with $M_p$ being the nucleon mass, $q^2 = -Q^2$ is the four-momentum
squared transferred from the neutrino to the $W^\pm$ boson of mass
$M_W$, and $G_F$ is the Fermi coupling. The Bjorken variable is given
by
\be
x = \frac{Q^2}{2qP} = \frac{Q^2}{W^2+Q^2-M_p^2} \cong \frac{Q^2}{W^2},
\label{6}
\ee
where the approximate equality in (\ref{6}) is valid in the relevant
range of $x \ll 0.1$. The fraction of the energy transfer from the
neutrino to the $W^\pm$ boson, $y$, is given by
\be
y = \frac{Q^2}{2M_pEx} \cong \frac{W^2}{s}.
\label{7}
\ee

For the subsequent discussion, it will be useful to replace the integration
over $dx$ in (\ref{2}) by an integration over $W^2$, rewriting (\ref{2}) as
\be
\sigma_{\nu N} (E) = \frac{G^2_F}{2 \pi} \int^{s-M_p^2}_{Q^2_{min.}} dQ^2
\left( \frac{M^2_W}{Q^2 + M^2_W} \right)^2 \int^{s-Q^2}_{M_p^2} 
\frac{dW^2}{W^2} \sigma_r (x,Q^2).
\label{8}
\ee
Due to the vector-boson propagator, contributions to the total cross section
for $Q^2 \gg M^2_W$ are strongly suppressed, and with $W^2 \le s$ and $s$ in the
ultrahigh energy range, $s \gg M^2_W$,
we expect the cross section to dominantly originate
from $x \approx Q^2/W^2 \ll 0.1$. 

In what follows, we concentrate on the 
(dominant) contribution due to $F_2^\nu (x,Q^2)$ in (\ref{8}) according
to (\ref{4}).\footnote{The contribution due to $F_L^\nu (x,Q^2)$ turned
out to be less than 6 \%, compare the discussion in connection with Table 4
below.
The contribution from the structure
function $F_3 (x,Q^2)$ in (\ref{4}), that is due to valence-quark interactions,
can be ignored.}

For small values of $x \lsim 0.1$, DIS of electrons and neutrinos on nucleons,
in terms of, respectively, the $\gamma^*p$ and the $(W^\pm, Z^0) p$ forward
scattering amplitude, proceeds via scattering of long-lived massive 
hadronic fluctuations, $\gamma^*(Z^0) \to q \bar q$ and $W^- \to \bar u d$
etc., that undergo diffractive forward scattering on the 
nucleon (CDP) \cite{Lanz1}.

For the flavor-symmetric $(q \bar q) N$ interaction at $x \ll 0.1$,
the neutrino-nucleon structure function, $F_{2}^{\nu N} (x,Q^2)$, 
and the electromagnetic structure function, $F_{2}^{eN} (x,Q^2)$, 
are related by $(1/n_f) F_2^{\nu N} (x,Q^2) = (1/\sum_q Q^2_q)
F_2^{eN} (x,Q^2)$, or
\be
F_{2,L}^{\nu N} (x, Q^2) = \frac{n_f}{\sum^{n_f}_q Q^2_q} F^{eN}_{2,L}
(x,Q^2),
\label{9}
\ee
where $n_f$ denotes the number of actively contributing quark flavors, and
$Q_q$ the quark charge, and $n_f/\sum_q Q^2_q = 18/5$ for $n_f = 4$ flavors of
quarks. As a consequence of the
proportionality (\ref{9}), the total neutrino-nucleon cross
section (\ref{8}) may be predicted by inserting the electromagnetic
structure function into (\ref{4}).

The electromagnetic structure function, 
$F^{ep}_2 (x,Q^2)$, is related to the total photoabsorption cross section,
$\sigma_{\gamma^*p} (W^2,Q^2)$, by\footnote{The low-$x$ approximation is used
for the factor in front of $\sigma_{\gamma^*p} (W^2,Q^2)$ in (\ref{10}).}
\be
F^{ep}_2 (x,Q^2) = \frac{Q^2}{4 \pi^2 \alpha} \sigma_{\gamma^*p} (W^2,Q^2).
\label{10}
\ee
In the CDP, as a consequence \cite{Lanz2, Lanz3}
of the interaction of the color dipole with
the gluon field in the nucleon, the photoabsorption cross section becomes
a function of the low-$x$ scaling variable, $\eta (W^2,Q^2)$,
\be
\sigma_{\gamma^*p} (W^2,Q^2) = \sigma_{\gamma^*p} (\eta(W^2,Q^2)) \sim
\sigma^{(\infty)} \left\{ \begin{array}{l@{\quad,\quad}l}
ln \frac{1}{\eta (W^2,Q^2)} & {\rm for}~ \eta(W^2,Q^2) \ll 1,\\
\frac{1}{2 \eta (W^2,Q^2)} & {\rm for}~ \eta(W^2,Q^2) \gg 1,
\end{array} \right.
\label{11}
\ee
where the cross section $\sigma^{(\infty)} \equiv \sigma^{(\infty)}(W^2)$ 
is of hadronic size, and,
at most, it depends weakly on $W^2$. Both, the dependence on the single
variable $\eta (W^2,Q^2)$ (for $\sigma^{(\infty)} \cong const.$) in 
(\ref{11}), and the specific functional form of this dependence,
are general consequences \cite{Lanz2, Lanz3}
of the color-gauge-invariant interaction of
a $(q \bar q)$ dipole with the
color field in the nucleon. Any specific ansatz for a parameterization of
the dipole-nucleon cross section has to provide an interpolation between the 
$ln (1/\eta (W^2,Q^2))$ and the $1/2 \eta (W^2,Q^2)$ dependence in
(\ref{11}). 
\begin{figure}[h]
\begin{center}
\hspace{-1cm}\epsfig{file=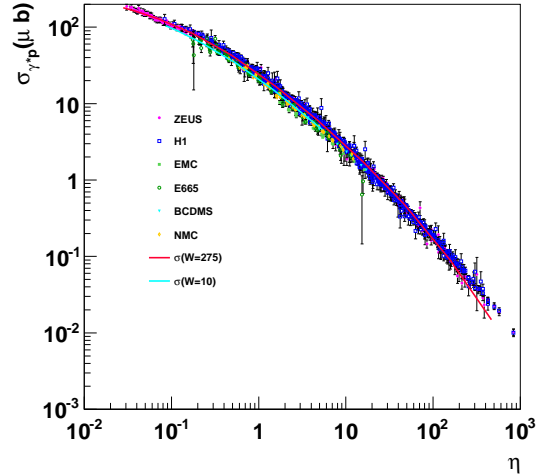,width=8cm} 
\vspace*{-1.2cm}  
\caption{{
The theoretical prediction \cite{Lanz2, Lanz3}
for the photoabsorption cross section 
$\sigma_{\gamma^* p} (\eta (W^2, Q^2))$ compared with the
experimental data on DIS.}}
\end{center}
\end{figure}
It is well known \cite{Lanz2}, compare Fig. 2, that the dependence (\ref{11})
on the single variable $\eta (W^2,Q^2)$ is fulfilled by the experimental
data with $\sigma^{(\infty)} \cong const.$ in the HERA energy range.
The saturation scale is given by \cite{Lanz2,1108, Lanz3}
\be
\Lambda^2_{sat} (W^2) = C_1 \left( \frac{W^2}{1 {\rm GeV}^2} \right)^{C_2},~~~
C_1= 0.34~GeV^2,~~~C_2 \cong 0.29.
\label{12}
\ee
The value of the exponent $C_2 \cong 0.29$ is fixed \cite{Lanz3} by
requiring consistency of the CDP with the pQCD-improved parton model.

We return to neutrino scattering. Employing relation (\ref{9}), we replace
the neutrino structure function, $F^\nu_2 (x,Q^2)$, in (\ref{4}) by the
electromagnetic one, $F^{ep}_2 (x, Q^2)$, or rather by the photoabsorption
cross section, compare (\ref{10}). The neutrino-nucleon total cross section
(\ref{8}) becomes\footnote{We restrict ourselves to the dominant term
$F^\nu_2 (x,Q^2)$ in (\ref{4}), ignoring $F_L (x,Q^2)$ and $F_3 (x,Q^2)$.}
\bqa
\sigma_{\nu N}(E) &= \frac{G^2_F M^4_W}{8 \pi^3 \alpha} 
\frac{n_f}{\sum_q Q^2_q} \int^{s-M_p^2}_{Q^2_{Min}} d Q^2 
\frac{Q^2}{(Q^2 + M^2_W)^2}
\nonumber \\
&\times \int^{s-Q^2}_{M_p^2} \frac{dW^2}{W^2} \frac{1}{2} (1 + (1-y)^2) 
\sigma_{\gamma^*p} (\eta (W^2, Q^2)).
\label{13}
\eqa

We first of all look at the ratio
\be
r(E) = \frac{\sigma_{\nu N}(E)_{\eta (W^2, Q^2) < 1}}
{\sigma_{\nu N} (E)}.
\label{14}
\ee
In (\ref{14}), $\sigma_{\nu N}(E)_{\eta (W^2, Q^2) < 1}$ denotes that
part of the total neutrino-nucleon cross section in (\ref{13})
that originates from
contributions from the saturation region of $\eta (W^2,Q^2) < 1$ in Fig. 1.
This part of the total cross section (\ref{13}) is obtained by imposing 
the cut of
$\eta(W^2,Q^2) < 1$ on the $(Q^2,W^2)$
integration domain in (\ref{13}). According to
(\ref{1}) and (\ref{12}), the restriction of $\eta (W^2,Q^2) < 1$ 
(for $Q^2_{Max.}\ge Q^2 \ge Q^2_{Min}= \Lambda_{sat}^2(M_p^2)-m_0^2$, and
$Q^2_{Max} \gg m^2_0$) 
upon employing $W_{Max}^2=s-Q^2$, yields
\bqa
  &&   W^2\ge W^2(Q^2)_{Min}
  = \left( \frac{Q^2+m^2_0}{C_1} \right)^{\frac{1}{C_2}},
    \nonumber \\
  &&   Q^2\le Q^2_{Max}=\Lambda_{sat}^2(s)
      \Bigl(1-C_2{{\Lambda^2_{sat}(s)}\over s}+
      o({{\Lambda^4_{sat}(s)}\over {s^2}})\Bigr).
\label{15}
\eqa
From (\ref{15}), for the ultrahigh-energy corresponding to $s = 10^{14}
{\rm GeV}^2$, with (\ref{12}), one finds $Q^2 < Q^2_{Max} = \Lambda^2_{sat} (s)
= 3.9 \times 10^3 {\rm GeV}^2
\ll s$. We observe that even for $s = 10^{14} {\rm GeV}^2$, the range of 
$Q^2 < Q^2_{Max}$ covered under restriction (\ref{15}) is smaller than
the $W^\pm$ mass squared, $M^2_W \approx 6.4 \times 10^3 {\rm GeV}^2$, that
determines the maximum of the $Q^2$-dependent factor in (\ref{13}).
We accordingly expect a small value of $r (E) \ll 1$.

The ratio $r(E)$ in (\ref{14}) is evaluated in two steps. In a first step,
we only rely on the very general low-x scaling restrictions for
$\sigma_{\gamma^*p} (\eta (W^2,Q^2))$ in (\ref{11}) with (\ref{12})
and derive an upper bound on $r(E) < \bar{r} (E)$ on $r(E)$. 
In a second step, 
we introduce a concrete representation for 
$\sigma_{\gamma^*p} (\eta (W^2,Q^2))$ in the CDP
that smoothly interpolates the regions
of $\eta (W^2,Q^2) < 1$ and $\eta (W^2, Q^2) > 1$ in (\ref{11}).

The ratio $r(E)$ in (\ref{14}), upon substituting (\ref{13}) and
taking into account (\ref{15}), becomes
\be
r(E) = \frac{\int^{Q^2_{Max}(s)}_{Q^2_{Min}} dQ^2 
\frac{Q^2}{(Q^2+ M^2_W)^2} \int^{s-Q^2}_{W^2(Q^2)_{Min}} \frac{dW^2}{W^2}
(1+(1-y)^2) \sigma_{\gamma^*p} (\eta(W^2,Q^2))}
{\int^{s-M_p^2}_{Q^2_{Min}} dQ^2 \frac{Q^2}{(Q^2+M^2_W)^2} \int^{s-Q^2}_{M_p^2}
\frac{dW^2}{W^2} (1 + (1-y)^2) \sigma_{\gamma^*p} (\eta (W^2,Q^2))}.
\label{16}
\ee
Using the scaling behaviour (\ref{11}) for $\eta (W^2,Q^2) < 1$ and
$\eta (W^2,Q^2) > 1$, we derive an upper limit, 
\bq
   r(E) < \bar{r} (E),
\label{17}
\eq
on the ratio $r(E)$ in (\ref{16}). Appropriately substituting the behaviour
(\ref{11}) of $\sigma_{\gamma^*p} (\eta (W^2,Q^2))$ into (\ref{16}),
and simplifying by putting $y=0$ in the numerator and $y=1$ in the
denominator, an upper bound on $r(E)$ reads\footnote{In the denominator
of (\ref{18}), we inserted the $1/2 \eta (W^2,Q^2)$ dependence only
valid for $\eta (W^2,Q^2) > 1$. We explicitly checked that the 
enlargement of the cross section as a consequence of this approximation
amounts to only a few percent in the energy range up to $E \sim 10^{14} {\rm GeV}$
under consideration.}
\be
\bar{r}(E) = \frac{2 \int^{Q^2_{Max}(s)}_{Q^2_{Min}} dQ^2 
\frac{Q^2}{(Q^2+ M^2_W)^2} \int^{s-Q^2}_{W^2(Q^2)_{Min}} \frac{dW^2}{W^2}
\ln \frac{1}{\eta (W^2,Q^2)}}
{\int^{s-M_p^2}_{Q^2_{Min}} dQ^2 \frac{Q^2}{(Q^2+M^2_W)^2} \int^{s-Q^2}_{M_p^2}
\frac{dW^2}{W^2} \frac{1}{2 \eta (W^2,Q^2)}}.
\label{18}
\ee
For $\Lambda^2_{sat} (s) < M^2_W \ll s$,
one finds that the numerator in (\ref{18}) is approximately given by
\be
  N(E) = \frac{1}{2} \frac{1}{2C_2} 
     \left( \frac{\Lambda^2_{sat} (s)}{M^2_W}\right)^2 
      +o(\left(\frac{\Lambda^2_{sat} (s)}{M^2_W}\right)^3 ).
\label{19}
\ee
The denominator in (\ref{18}) becomes
\be
  D(E) =  \frac{1}{2C_2} \left(\frac{\Lambda^2_{sat} (s)}{M^2_W}\right)
    \left(1+o({{M_W^2}\over s}\log{{M_W^2}\over s})\right).
\label{20}
\ee
Inserting (\ref{19}) and (\ref{20}) into (\ref{18}), we find the upper
bound on $r(E)$,
\be
r(E) < \bar{r} (E) = \frac{1}{2} 
\frac{\Lambda^2_{sat} (s)}{M^2_W}.
\label{21}
\ee
Numerical values of $\bar r(E)$, using (\ref{12}), are given in 
Table 1, together with the results
for $r(E)$ resulting from an explicit expression for $\sigma_{\gamma^*p}
(\eta (W^2,Q^2))$ from the CDP to be discussed below.

\begin{table}
\begin{center}
\begin{tabular}{|l|c|c|c|}
\hline
$E ({\rm GeV})$ & $\bar{r} (E)$ & $r(E)\vert_{\rm Table 3}$ & 
$r(E) \vert_{\rm Table 4}$\\
\hline
$10^6$ & $ 1.74 \times 10^{-3}$ & $ 1.40 \times 10^{-3}$ 
& $4.58 \times 10^{-3}$\\
\hline
$10^{10}$ & $ 2.51 \times 10^{-2}$ & $1.63 \times 10^{-2}$ 
& $2.55 \times 10^{-2}$\\
\hline
$10^{14}$ & $ 3.63 \times 10^{-1}$ & $1.76 \times 10^{-1}$ 
& $ 1.96 \times 10^{-1}$\\
\hline
\end{tabular}
\end{center}
\caption{The upper bound, $\bar r(E) > r(E)$, on the fraction of the total
neutrino-nucleon cross section originating from the saturation region
of\break $\eta (W^2,Q^2)<1$. The results for $\bar r(E)$ 
in the second column
are based on (\ref{21}) with (\ref{12}).
The results for $r(E)\vert_{\rm Table~ 3}$ are 
based on evaluating (\ref{16}) upon substitution of (\ref{22})
with (\ref{25}). The results for $r(E) \vert_{\rm Table~4}$ are
based on evaluating (\ref{16}) upon substitution of (\ref{29}) with
(\ref{25}).}
\end{table}
According to (\ref{21}) and Table 1, the fraction of the total 
neutrino-nucleon cross section arising from the saturation region is
strongly suppressed. The saturation region contributes less than a
few percent, except for extremely ultrahigh energies of order
$E \simeq 10^{14} {\rm GeV}$.

We turn to an evaluation of the neutrino-nucleon cross section based
on an explicit form of $\sigma_{\gamma^*p} (\eta (W^2,Q^2))$ in the CDP.

The CDP leads to a remarkably simple form of the photoabsorption cross
section that moreover can be represented by a 
closed expression,\footnote{We note
that the closed form for the photoabsorption cross section in (\ref{22})
with (\ref{23}) contains the simplifying assumption of ``helicity independence''
leading to $F^{ep}_L = 0.33~~F^{ep}_2$ rather than $F^{ep}_L = 0.27~~F^{ep}_2$.
This simplifying approximation is unimportant in the present context. Compare
refs. \cite{Lanz3, Ku-Schi} for the refinement that implies the result
$F^{ep}_L = 0.27~~F^{ep}_2$
that is consistent with the HERA experimental observations.} \cite{Lanz2, Lanz3}
\bqa
\sigma_{\gamma^* p} (W^2, Q^2)& = & 
\sigma_{\gamma^* p} (\eta (W^2, Q^2)) + O \left( \frac{m^2_0}{\Lambda^2_{\rm
      sat} (W^2)} \right) =  \nonumber \\ 
& = & \frac{\alpha R_{e^+ e^-}}{3\pi} \sigma^{(\infty)} 
(W^2) I_0 (\eta (W^2,Q^2)) + O \left(
  \frac{m^2_0}{\Lambda^2_{\rm sat} (W^2)} \right) , 
\label{22}
\eqa
where
\bqa
I_0 (\eta (W^2, Q^2)) & = & 
\frac{1}{\sqrt{1 + 4\eta (W^2, Q^2)}} \ln \frac{\sqrt{1 + 4 \eta (W^2, Q^2)}
  +1}{\sqrt{1+4\eta(W^2, Q^2)}-1} \cong  \label{23}  \\
& \cong& \left\{  \matrix{  \ln \frac{1}{\eta(W^2, Q^2)} + O (\eta \ln \eta ), 
~~{\rm for}~ \eta (W^2, Q^2) \rightarrow \frac{m^2_0}{\Lambda^2_{\rm sat}
  (W^2)}, \cr
\frac{1}{2\eta (W^2, Q^2)} + O \left( \frac{1}{\eta^2}\right) , ~ {\rm
  for}~ \eta (W^2 , Q^2) \rightarrow \infty ,  } \right. \nonumber
\eqa
and 
\be
R_{e^+ e^-} = 3 \sum_q Q^2_q .
\label{24}
\ee
Comparing (\ref{22}) and (\ref{23}) with (\ref{11}), one notes that  
(\ref{22}) smoothly
interpolates the regions of $\eta(W^2,Q^2) \ll 1$ and $\eta (W^2,Q^2) \gg 1$
in (\ref{11}).

The (weak) energy dependence of the dipole cross section
$\sigma^{(\infty)} (W^2)$ in (\ref{22}) is determined by consistency
of $\sigma_{\gamma^*p} (W^2,Q^2)$ with Regge behavior \cite{Lanz2, Donnachie} 
in the
photoproduction limit of $\sigma_{\gamma p} (W^2) = \sigma_{\gamma^*p}
(W^2,Q^2 = 0)$, and alternatively, by consistency with the double-logarithmic
fit to photoproduction by the Particle Data Group,
\be
\sigma^{(\infty)} (W^2) = \frac{3\pi}{R_{e^+e^-}\alpha}
\frac{1}{\ln \frac{\Lambda^2_{sat} (W^2)}{m^2_0}} 
\left\{ \begin{array}{l@{\quad\quad}l}
\sigma^{Regge}_{\gamma p} (W^2), & ~ \\
&~ \\
\sigma^{PDG}_{\gamma p} (W^2). & ~
\end{array} \right.
\label{25}
\ee
The fits to photoproduction, compare refs. \cite{Lanz2}, \cite{Donnachie} and
\cite{PDG}  (in units of $mb$,
with $W^2$ in ${\rm GeV}^2$) are explicitly given by
\bqa
\sigma^{(a)}_{\gamma p} (W^2) & = & 0.0635 (W^2)^{0.097} + 0.145 (W^2)^{-0.5},
\label{26} \\
\sigma^{(b)}_{\gamma p} (W^2) & = & 0.0677 (W^2)^{0.0808} 
+ 0.129 (W^2)^{-0.4525} \nonumber \\
\sigma^{(c)}_{\gamma p} (W^2) & = & 0.003056 \left(33.71 + \frac{\pi}{M^2}
\ln^2 \frac{W^2}{(M_p + M)^2} \right) \nonumber \\
& + & 0.0128 \left( \frac{(M_p+M)^2}{W^2}
\right)^{0.462},\nonumber
\eqa
where $M_p$ stands for the proton mass and $M = 2.15 {\rm GeV}$. 
Concerning
the energy dependence of the photoabsorption cross section in (\ref{22}), we
note that the growth $\sigma_{\gamma^*p}(W^2,Q^2) \sim (\ln W^2)
(W^2)^{C_2}$ in the color-transparency region (for 
$\sigma^{(\infty)} (W^2) \sim \sigma^{PDG}_{\gamma p}(W^2)
/\ln{{\Lambda^2_{sat}(W^2)}\over{m_0^2}} $) of 
$\eta (W^2,Q^2) > 1$ turns into the slower growth of
$\sigma_{\gamma^*p} (W^2,Q^2) \sim (\ln W^2)^2$, once the
saturation limit of $\eta (W^2,Q^2) < 1$ is reached.

In Table 2, we present the results for the neutrino-nucleon cross section
based on (\ref{13})\footnote{The CDP contains the limit of $Q^2\to 0$, 
such that $Q^2_{Min}$ may be put to $Q_{Min}^2=0$ in (\ref{13}).  
The actual dependence on $Q^2_{Min}$ is negligible, as long as 
{\boldmath$0$}$\lsim Q^2_{Min}\lsim M_p^2$.  We also note tht the replacement of the 
lower limit $W^2\ge M_p^2$ by $W^2\ge {\rm const}~ M_p^2$
for e.g. const$\le 20$ leads to an insignificant change of the
neutrino cross section.}
 upon substitution of the photoabsorption cross section from
(\ref{22}) with $\Lambda^2_{sat} (W^2)$ from (\ref{12}), $m^2_0 = 0.15 
{\rm GeV}^2$, and
$\sigma^{(\infty)} (W^2)$ determined by
(\ref{25}) and (\ref{26}). The results in Table 2 for $\sigma^{(b)}_{\nu N}
(E)$ and $\sigma^{(c)}_{\nu N} (E)$ based on $\sigma^{(\infty)} (W^2)$ from
the Regge fit (b) and the PDG fit (c), respectively, coincide in good
approximation. The enhancement of the cross section $\sigma^{(a)}_{\nu N} (E)$
relative to $\sigma^{(b,c)}_{\nu N} (E)$ is a consequence of the stronger
increase of the Pomeron contribution ($(W^2)^{0.097}$ versus $(W^2)^{0.0808}$)
in $\sigma^{(\infty)} (W^2)$ originating from (\ref{26}). At the highest
energy under consideration, $E = 10^{14} {\rm GeV}$, the enhancement reaches a
factor of about 1.5. Concerning the energy dependence, by comparing 
neighboring results in Table 2 for $E \ge 10^8 {\rm GeV}$, one notes an increase
(only) slightly stronger than expected from the proportionality to
$\Lambda^2_{sat} (s) \sim s^{C_2}$ in the estimate (\ref{20}). This is a
consequence of the energy dependence (\ref{25})
of $\sigma^{(\infty)} = \sigma^{(\infty)}
(W^2)$ ignored in (\ref{20}).

\begin{table}
\begin{tabular}{| r|| r | r | r | r | r| r |}\hline
 $E$ & 1.0E+04 & 1.0E+06 & 1.0E+08 & 1.0E+10 & 1.0E+12 & 1.0E+14 \cr
\hline
 $\sigma^{(a)}_{\nu N}$ & 1.28E-34 & 1.91E-33 & 1.09E-32 & 5.36E-32 & 2.60E-31
      & 1.23E-30 \cr
\hline
 $\sigma^{(b)}_{\nu N}$ & 1.21E-34 & 1.68E-33 & 8.96E-33 & 4.11E-32 & 1.85E-31
      & 8.15E-31 \cr     
\hline
 $\sigma^{(c)}_{\nu N}$ & 1.19E-34 & 1.69E-33 & 9.26E-33 & 4.29E-32 & 1.88E-31
      & 7.77E-31 \cr
\hline
\end{tabular}

\caption{The prediction of the neutrino-nucleon cross section,
$\sigma_{\nu N}^{(a,b,c)} [cm^2]$, from the CDP as a function of
the neutrino energy, $E [{\rm GeV}]$.
Compare text for details.}
\end{table}

We return to the question of the relative contribution to the neutrino
cross section from the saturation region relative to the color-transparency
region. We subdivide the neutrino cross section into the sum 
\be
\sigma^{(c)}_{\nu N} (E) = \sigma^{(c)}_{\nu N} (E)_{\eta(W^2,Q^2)< 1} +
\sigma^{(c)}_{\nu N} (E)_{\eta(W^2,Q^2) > 1}.
\label{27}
\ee
The results are shown in Table 3.
\begin{table}[h]
\begin{tabular}{| r|| r | r | r | r | r| r |}\hline
 $E$ & 1.0E+04 & 1.0E+06 & 1.0E+08 & 1.0E+10 & 1.0E+12 & 1.0E+14 \cr
\hline
 $\sigma^{(c)}_{\nu N}$ & 1.19E-34 & 1.69E-33 & 9.26E-33 & 4.29E-32 & 1.88E-31
           & 7.77E-31 \cr
\hline
  $\eta>1$ &1.19E-34 & 1.68E-33 & 9.22E-33 & 4.22E-32 & 1.77E-31 & 6.41E-31\cr
\hline
  $\eta<1$ &1.14E-37 & 2.37E-36 & 4.15E-35  & 6.97E-34 & 1.08E-32 & 1.37E-31 \cr
\hline
\end{tabular}
\caption{The contributions to the neutrino-nucleon cross section 
$\sigma^{(c)}_{\nu N} (E) ~ [cm^2]$  as a function of
$E[GeV]$
from the
color transparency $(\eta (W^2,Q^2) > 1)$
and the saturation $(\eta (W^2,Q^2) < 1)$ region compared with the full
cross section, $\sigma^{(c)}_{\nu N} (E)$ {taken from Table 2}.}
\end{table}
From Table 3, one finds that the fraction of the total cross section
originating from the saturation region, $r(E)$ in (\ref{14}) and (\ref{16}), 
increases
from $r(E=10^6 {\rm GeV})\vert_{Table 3} \cong 1.40 \cdot 10^{-3}$ to 
$r(E=10^{14} {\rm GeV})\vert_{Table 3} \cong 1.76 \cdot 10^{-1}$.
The increase is consistent with the upper bound (\ref{21}), compare Table 1. 
With increasing energy, there is a strong increase
from the saturation region, but even at $E = 10^{14} {\rm GeV}$ its contribution
is of the order of only 17\%. 

The result that the dominant part of the neutrino-nucleon cross section is
due to contributions from large values of $\eta (W^2,Q^2) \gg 1$ requires
further examination. For e.g. a value of $Q^2 = 10^4 {\rm GeV}^2 \cong M^2_W$,
and for $W^2$ below $W^2 \le 10^5 {\rm GeV}^2$ (or $x \le 0.1$), one finds that
$\eta (W^2,Q^2)$ reaches values of $\eta (W^2, Q^2) \le \eta_{Max} (W^2,Q^2)
\cong 10^3$. For such large values of $\eta (W^2,Q^2)$, as previously
analysed \cite{Lanz2, Lanz3}, the theoretical expression (\ref{22}) for
the photoabsorption cross section must be corrected by elimination of
contributions from high-mass $(q \bar q)$ fluctuations,
$\gamma^* \to q \bar q$, of mass $M_{q \bar q}$. The life time of 
high-mass fluctuations in the rest frame of the nucleon becomes too short
to be able to actively contribute to the $q \bar q$-color-dipole 
interaction. The
restriction on the $q \bar q$ mass, $m^2_0 \le M^2_{q \bar q} \le m^2_1 (W^2)$
is taken care of by the energy-dependent upper bound, $m^2_1 (W^2)$, where
\be
m^2_1 (W^2) = \xi \Lambda^2_{sat} (W^2),
\label{28}
\ee
and empirically $\xi = 130$ \cite{Lanz3}. Employing the restriction 
(\ref{28}) extends the validity of the CDP to high values of 
$\eta (W^2,Q^2) \gg 1$.

Explicitly, one finds that (\ref{22}) must be modified by a factor that
depends on the ratio of $\xi/\eta (W^2,Q^2)$. One obtains \cite{Lanz3}
\bqa
\sigma_{\gamma^*p} (W^2,Q^2) & = & \frac{\alpha R_{e^+e^-}}{3 \pi}
\sigma^{(\infty)} (W^2) I_0 (\eta (W^2,Q^2)) \nonumber \\
& \times & \frac{1}{3} \left( G_L \left( \frac{\xi}{\eta (W^2,Q^2)} \right)
+ 2 G_T \left( \frac{\xi}{\eta (W^2,Q^2)} \right) \right) \nonumber \\
& + & O \left(
  \frac{m^2_0}{\Lambda^2_{\rm sat} (W^2)} \right)
\label{29}
\eqa
where
\bqa
&& \frac{1}{3} \left( G_L \left( \frac{\xi}{\eta (W^2,Q^2)} \right) + 2 G_T
\left( \frac{\xi}{\eta (W^2,Q^2)} \right) \right) = \nonumber \\
&& \frac{1}{\left( 1 + \frac{\xi}{\eta (W^2,Q^2)} \right)^3}
\left( \left( \frac{\xi}{\eta (W^2,Q^2)} \right)^3 + 2 \left( 
\frac{\xi}{\eta (W^2,Q^2)} \right)^2 + \left( \frac{\xi}{\eta(W^2,Q^2)} \right)
\right) \nonumber \\
&& \cong \left\{ \begin{array}{l@{\quad,\quad}l}
1 & {\rm for}~\eta (W^2,Q^2) \ll \xi = 130 \\
\frac{\xi}{\eta (W^2,Q^2)} & {\rm for}~\eta (W^2,Q^2) \gg \xi = 130
\end{array} \right.  ;
\label{30}
\eqa
We note in passing that the theoretical prediction shown in Fig. 2
includes\cite{Lanz3} the large-$\eta (W^2,Q^2)$ correction 
(\ref{29})\footnote{The photoabsorption cross section obtained from the
simple closed expression (\ref{29}) coincides within a (negative) deviation of
up to approximately 25 \% with the results shown in fig. 2 that are based on
the more elaborate treatment in ref.\cite{Lanz3}, compare footnote 8} .

\newcommand{\rb}[1]{\raisebox{0.7ex}[-0.7ex]{#1}}
\begin{table}[h]
\begin{tabular}{|l|l|l|l|l|l|l|}\hline
E & 1.0E+04 & 1.0E+06 & 1.0E+08 & 1.0E+10 & 1.0E+12 & 1.0E+14 \\
\hline
 & 1.19E-34 & 1.69E-33 & 9.26E-33
& 4.29E-32 & 1.88E-31 & 7.77E-31 \\
\cline{2-7}
\rb{$\sigma^{(c)}_{\nu N}$} & 3.85E-35 &5.15E-34 &4.17E-33 &2.73E-32 
&1.49E-31 &6.96E-31 \\
\cline{2-7}
 & 3.19E-35 & 3.80E-34 & 2.83E-33 & 1.75E-32 &9.12E-32 & 4.11E-31 \\
\hline
\end{tabular}
\caption{The neutrino-nucleon cross section, 
$\sigma^{(c)}_{\nu N} (E) [cm^2]$,
as a function of the neutrino energy $E [GeV]$
upon imposing the restriction
(\ref{28}) on the mass of actively contributing $q \bar q$ fluctuations 
(3rd and 4th line) compared with the result from Table 3 (2nd line) that ignores
the restriction (\ref{28}). The results in the 3rd and 4th line are based
on $\Lambda^2_{sat} (W^2) \sim (W^2)^{C_2}$ with $C_2 = 0.29$ and $C_2 = 0.27$,
respectively.}
\end{table}


In Table 4, third and fourth line,
we present our final results for the neutrino-nucleon cross
section based on substituting (\ref{29}) into (\ref{13}). The PDG result
for $\sigma^{(\infty)} (W^2)$ in (\ref{25}) is used, and, for comparison, the
result for $\sigma^{(c)}_{\nu N} (E)$
from Table 2 ( i.e. 
$\sigma^{(c)}_{\nu N} (E)$  without the 
restriction (\ref{28}))
is again shown in the second line of Table 4. 
We explicitly verified that the addition in (\ref{13}) of the contribution
corresponding to the longitudinal structure function according to (\ref{4})
diminishes the neutrino cross section in Table 4 by less than 6 \% in the
whole range of neutrino energies under consideration.
In order to demonstrate the
sensitivity under variation of the exponent $C_2$ of the energy dependence
of the saturation scale, 
$\Lambda^2_{sat} (W^2) \sim (W^2)^{C_2}$, in Table 4,
we give the neutrino-nucleon cross section for $C_2 = 0.29$ and $C_2 = 0.27$.
Both values are consistent with the available experimental information on DIS.

\begin{figure}[h!]
\vspace*{-0.7cm}
\begin{center}
\hspace{-1cm}\epsfig{file=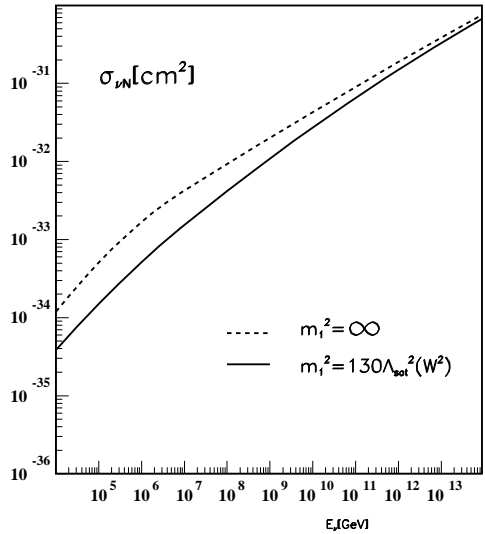,width=8.5cm} 
\vspace*{-0.2cm}  
\caption{The effect on the neutrino-nucleon cross section of 
excluding inactive high-mass $q \bar q$ fluctuations.}
\end{center}
\end{figure}

The results from Table 4 (2nd and 3rd line) are 
graphically represented in Fig. 3.
With increasing neutrino energy, the exclusion of inactive large-mass
$q \bar q$ fluctuations by the restriction of $M^2_{q \bar q} < m^2_1 (W^2)
= \xi \Lambda^2_{sat} (W^2)$, where $\xi = 130$, becomes less important.
Most of the contributions to the neutrino-nucleon cross section in the extreme
ultrahigh-energy limit ($E \simeq 10^{14}$ GeV) are due to moderately large
values of $\eta (W^2,Q^2)$ that correspond to $q \bar q$ fluctuations of
sufficiently long life time. Quantitatively, from Table 4, at $E = 10^4$ GeV
the cross section is diminished by a factor of 0.32, while at $E = 10^{14}$
GeV, this factor is equal to 0.89. This effect is also seen in the ratio
$r(E)$ in Table 1. At $E = 10^6$ GeV, the ratio $r(E)$ exceeds the crude
estimate of $\bar r (E)$ from (\ref{18}).

\begin{figure}[h!]
\begin{center}
\hspace{-1cm}\epsfig{file=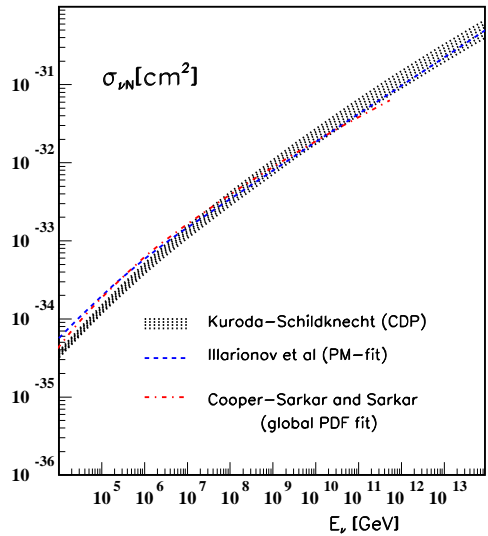,width=8.5cm} 
\vspace*{-0.2cm}  
\caption{Comparison of the CDP prediction for the neutrino-nucleon
cross section, $\sigma_{\nu N} (E) [cm^2]$, according to (\ref{13})
with (\ref{29}) and $\sigma^{PDG}_{\gamma p} (W^2)$  from (\ref{25}),
with the predictions from the pQCD-improved parton model. The band of the
prediction from the CDP illustrates the sensitivity of $\sigma_{\nu N} (E)$
under variation of the exponent 
$C_2$ in $\Lambda^2_{sat} (W^2) \sim (W^2)^{C_2}$
between $C_2 = 0.27$ and $C_2 = 0.29.$
}
\end{center}
\end{figure}

In Fig. 4, we compare our final results 
for the neutrino-nucleon
cross section, $\sigma_{\nu N} (E) \equiv \sigma_{\nu N}^{(c)}
(E)$  from Table 4, 3rd and 4th line,
based on the CDP, with the ones obtained \cite{a, i}
by employing the parton distributions from a conventional perturbative
QCD (pQCD) analysis of DIS. Fig. 4 shows consistency of our CDP results
with the ones from the pQCD-improved parton model. Our predictions are also
consistent with the ones in ref. \cite{h}.

A series of recent papers \cite{Block2} - \cite{Block4} treats DIS
at HERA energies and ultra-high-energy neutrino scattering by adopting
an ansatz with an $(\ln W^2)^2$ dependence
of the underlying hadron-nucleon cross section. 
The ansatz is based on the asymptotic behavior of strong-interaction
cross sections as $(\ln W^2)^2$ due to Heisenberg\cite{Heisenberg}
and Froissart\cite{Froissart}.

The ansatz of
$F^{ep}_2 (x,Q^2) \sim \sum_{n,m = 0,1,2} a_{nm} (\ln Q^2)^n
(\ln (1/x))^m$, with seven free fit parameters \cite{Block2} -
\cite{Block4},
yields a successful representation of the HERA experimental results for
all $x$ and $Q^2$ in the region of $ x \lsim 
0.1$. The subsequent evaluation \cite{Block2} - \cite{Block4} of the
neutrino-nucleon cross section with this ansatz for $F^{ep}_2
  (x,Q^2)$, essentially according to (\ref{9}) and (\ref{13}), for
$E \gsim 10^9 GeV$ led to a cross section that is 
suppressed relative to 
pQCD results, and, consequently, also in comparison with our CDP predictions.
Compare Fig. 5.

\begin{figure}[h!]
\vspace*{-0.7cm}
\begin{center}
\hspace{-1cm}\epsfig{file=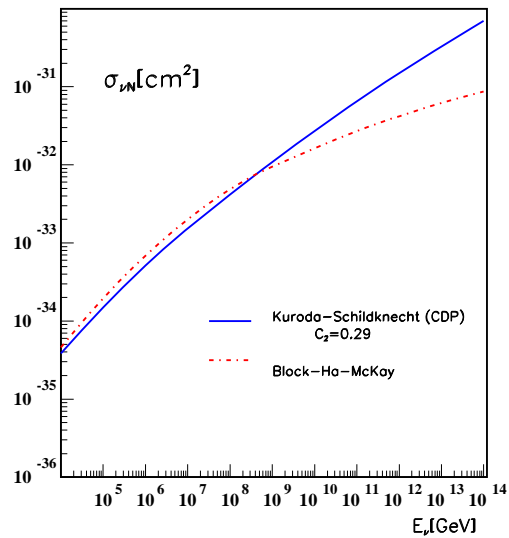,width=8.5cm} 
\vspace*{-0.2cm}  
\caption{A comparison of the results for the neutrino-nucleon
cross section from the CDP according to fig. 4 with the results
from the ``Froissart-inspired'' ansatz from \cite{Block3}.}
\end{center}
\end{figure}

Since the CDP contains an $(\ln W^2)^2$ dependence, compare e.g.
the discussion immediately following (\ref{26}), the result of Fig. 5 may look
like an inconsistency. The apparent inconsistency is resolved in fig. 6.
Figure 6 shows the prediction for the neutrino-nucleon cross section from the
CDP for an extended energy range up to $E = 10^{24} GeV$. As seen in fig. 6,
in consistency with the $(\ln W^2)^2$ dependence of $\sigma_{\gamma^*p}
(W^2,Q^2)$ in the saturation region of $\eta (W^2,Q^2) < 1$, also the
CDP implies a decreasing growth of the neutrino-nucleon cross section. In
distinction from the prediction from the ``Froissart-inspired'' ansatz,
the decreasing growth of the cross section in the CDP is shifted to energies
above $E \cong 10^{14} GeV$.

\begin{figure}[h!]
\vspace*{-0.7cm}
\begin{center}
\hspace{-1cm}\epsfig{file=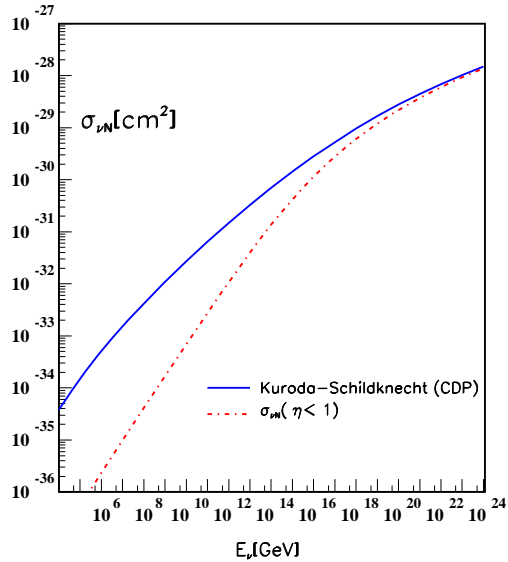,width=8.5cm} 
\vspace*{-0.2cm}  
\caption{The neutrino-nucleon total cross section, $\sigma_{\nu N} (E)
\equiv \sigma_{\nu N}^{(c)} (E)$, from the CDP as a function of the neutrino
energy $E$ for the extended range of energies up to $E = 10^{24} GeV$. For
comparison, we also show that part of the cross section, $\sigma_{\nu N}
(E) \vert_{\eta (W^2,Q^2)< 1}$, that is obtained upon restricting the 
contributions of $\sigma_{\gamma^*p} (W^2,Q^2)$ to the neutrino-nucleon
cross section to the saturation region of $\eta (W^2,Q^2) < 1$.}
\end{center}
\end{figure}

In Fig. 6, we explicitly demonstrate that the reduced growth of the 
neutrino cross
section with increasing energy is directly connected with the 
increasingly smaller contribution due to 
$\sigma_{\nu N}^{(c)} 
(E)_{\eta(W^2,Q^2)>1}$ in (\ref{27}). 
In the ultra-ultra-high-energy limit,
the neutrino-nucleon cross section in (\ref{13})
becomes saturated by contributions from that region of the 
photoabsorption cross section where the
$(\ln(W^2))^2$ dependence becomes dominant.

We must conclude that the requirement of
a ``Froissart-like'' ansatz for the underlying
hadron-nucleon cross section {\it by itself does not} imply a weaker 
growth, compared with e.g. the pQCD prediction, for the neutrino-nucleon
cross section above $E = 10^9 GeV$. 
It is the combination of the energy dependence for
$F_2^{ep} (x,Q^2)$, contained in $\ln (1/x)$ and $(\ln (1/x))^2$ terms,
with the seven-free-parameter fit to the ad hoc polynomial $\ln Q^2$
dependence of the coefficients of the $\ln (1/x)$ and $(\ln (1/x))^2$
terms that leads to a suppression above $E=10^9$ GeV.

In the CDP, the $Q^2$ dependence is uniquely fixed by the $Q^2$
dependence of the ``photon-wave function'', i.e.  the transition of the
(virtual) photon to $q \bar q$ dipole states with subsequent propagation
of these $q \bar q$ states of mass $M_{q \bar q}$. The
interaction of the $q \bar q$ color dipoles is restricted by being a
gauge-invariant interaction with the gluon field in the nucleon. 

Taking
into account the more detailed dynamics of the CDP, and the much smaller
number of free fit parameters, compared with the $\ln (1/x)$ and
$(\ln (1/x))^2$ ansatz, we are thus led to disagree with the conclusion of
an  onset of a suppression of the neutrino-nucleon cross section
for $E \gsim 10^9 GeV$ implied by the analysis 
\cite{Block2} - \cite{Block4} of the ``Froissart-inspired''
ansatz.

A  suppression, in the sense of a reduced growth of the total
neutrino-nucleon cross section with increasing energy, is expected to
occur, however, for neutrino energies beyond $E = 10^{14} GeV$.

\bigskip

\centerline{\Large\bf Acknowledgement}

Questions on the subject matter by Paolo Castorina and by participants
of the Oberwoelz symposium on Quantum Chromodynamics, History and Prospects
(Oberwoelz, Austria, September 3 -- 8, 2012) are gratefully acknowledged.

\end{document}